\documentclass[
 reprint,
nofootinbib,
 amsmath,amssymb,
 aps,
]{revtex4-2}

\usepackage{graphicx}% Include figure files
\usepackage{dcolumn}% Align table columns on decimal point
\usepackage{bm}
\usepackage{xcolor}
\usepackage{subfigure}% bold math
\usepackage{hyperref}% add hypertext capabilities
\usepackage[mathlines]{lineno}% Enable numbering of text and display math
\usepackage{graphicx}
\usepackage{hyperref}

\begin{document}

\title{Tensor resonances in teleparallel Gauss–Bonnet branes}

\author{J. V. R. Alencar}
\affiliation{Department of Teleinformatics Engineering, Federal University of Ceará (UFC), Fortaleza, CE, 60440-900, Brazil.}

\author{A. R. P. Moreira}
\email{allan.moreira@fisica.ufc.br}
\affiliation{Secretaria da Educa\c{c}\~{a}o do Cear\'{a} (SEDUC), Coordenadoria Regional de Desenvolvimento da Educa\c{c}\~{a}o (CREDE 9),  Horizonte, Cear\'{a}, 62880-384, Brazil.}
\affiliation{Postgraduate Program in Electrical and Computer Engineering, Federal University of Cear\'{a}, Sobral, Cear\'{a}, 62010-560, Brazil.}

\author{F. C. E. Lima}
\email{cleiton.estevao@ufabc.edu.br}
\affiliation{Centro de Matématica, Computação e Cognição (CMCC), Universidade Federal do ABC (UFABC), Av. dos Estados 5001, CEP 09210-580, Santo André, São Paulo, Brazil.}

\author{J. B. R. Silva}
\affiliation{Department of Teleinformatics Engineering, Federal University of Ceará (UFC), Fortaleza, CE, 60440-900, Brazil.}

\begin{abstract}
We construct an analytical thick-brane solution in linear teleparallel Gauss-Bonnet gravity using a first-order formalism generated by a sine-Gordon superpotential. The resulting asymptotically $\mathrm{AdS}_5$ configurations exhibit brane splitting controlled by the dimensionless parameter $q=4\alpha k^2$. An analytical splitting condition is derived and combined with tensor-stability requirements to identify the physically viable parameter region. We show that the tensor spectrum is free of tachyonic instabilities and supports a normalizable graviton zero mode. The massive sector exhibits odd-parity gravitational resonances whose quasi-localization is significantly enhanced near the stability boundary. Our results establish a direct link between brane splitting, tensor stability, and gravitational resonances in teleparallel Gauss-Bonnet braneworlds.
\end{abstract}


\maketitle

\section{Introduction}

Extra-dimensional braneworld models provide a geometric framework for recovering four-dimensional gravity from a higher-dimensional spacetime. In the Randall--Sundrum scenario, gravity is localized by a warped extra dimension, while thick-brane extensions replace the idealized thin brane by smooth scalar-field configurations, allowing a more realistic description of the bulk geometry \cite{Randall1999,Randall1999b,DeWolfe2000,Gremm2000}. The recovery of four-dimensional gravity requires the localization of the massless tensor mode, whereas the massive Kaluza--Klein spectrum encodes information about the extra dimension. Since the brane structure is sensitive to both the matter content and the underlying gravitational theory, modified gravity models can significantly alter the warp factor, energy-density distribution, and graviton localization properties, motivating extensive investigations in curvature-based, teleparallel, symmetric teleparallel, and Gauss--Bonnet braneworld scenarios \cite{Chen2021MimeticFR,Andrade:2023rbe,Fu2021fQBrane,Moreira2021fTT,Yang2022BornInfeld,Xu2023GBMultikink,Silva2026ESGBBraneworld}.

The internal structure of thick branes plays a crucial role in determining the spectrum of gravitational perturbations. In several modified gravity frameworks, geometric corrections can generate brane splitting, multikink configurations, and multi-well effective potentials, leading to gravitational resonances while preserving the localization and stability of the tensor zero mode \cite{Chen2021MimeticFR,Xu2023GBMultikink,Silva2026ESGBBraneworld}. Besides curvature-based theories, teleparallel gravity provides an alternative geometric description in which gravity is encoded in torsion rather than curvature. Although the teleparallel equivalent of general relativity reproduces the same dynamics as general relativity, its modified extensions lead to distinct braneworld phenomenology. In particular, $f(T)$, $f(T,\mathcal{T})$, and $f(Q)$ theories, as well as Born--Infeld determinantal gravity, have been shown to modify the brane thickness, tensor effective potential, graviton localization, and massive Kaluza--Klein spectrum, while allowing stable analytical thick-brane solutions with rich internal structures \cite{Tan2021fTResonances,Moreira2021fTT,Fu2021fQBrane,Yang2022BornInfeld,Moreira:2021uod,Moreira2024fQT}.

The tensor sector provides a fundamental consistency test for braneworld models, since a viable configuration must ensure the localization of the graviton zero mode, the absence of tachyonic instabilities, and a well-defined perturbation spectrum. Recent studies have shown that teleparallel braneworlds satisfy these requirements and remain stable under linear perturbations, reinforcing the robustness of torsion-based gravity as a framework for extra-dimensional scenarios \cite{Zhao2025fTLinearPerturbations}. Beyond the zero mode, the massive Kaluza--Klein spectrum offers valuable information about the internal structure of the brane. In particular, geometric modifications can generate resonant graviton states, which appear as quasi-localized modes associated with peaks in the relative probability and long-lived quasinormal oscillations \cite{Tan2021fTResonances,Tan2024RastallQNMs,Tan2025DoubleBraneworld}. Similar phenomena have been reported in curvature-based and Gauss--Bonnet braneworlds, indicating that the spectrum of massive tensor modes provides a sensitive probe of the underlying brane geometry and its internal structure \cite{Chen2021MimeticFR,Xu2023GBMultikink}.

Teleparallel Gauss--Bonnet gravity, described by $f(T,T_G)$ theories, extends teleparallel gravity through the inclusion of the torsional Gauss--Bonnet invariant $T_G$, providing a natural framework to investigate higher-order geometric effects in braneworld scenarios \cite{Kofinas2014,Kofinas2014Cosmo,Bahamonde2016}. In five dimensions, these corrections can significantly influence the brane structure and its perturbative properties, while recent studies indicate that gravitational waves propagate consistently at the speed of light within this framework \cite{Mishra2025TeleparallelGBGW}. Although $f(T,T_G)$ gravity has been explored in the context of thick branes and fermion localization, revealing important modifications in effective potentials and localization mechanisms, a comprehensive analytical model simultaneously addressing brane splitting, tensor stability, graviton localization, and massive gravitational resonances is still lacking. The present work aims to fill this gap.

In this letter, we construct an analytical thick-brane solution in linear teleparallel Gauss--Bonnet gravity, $f(T,T_G)=-T+\alpha T_G$, by employing a first-order formalism driven by a sine-Gordon superpotential. The resulting configurations are regular, asymptotically $AdS(_5)$, and exhibit a controllable internal structure governed by the dimensionless parameter ($q=4\alpha k^2$). An analytical condition for brane splitting is derived from the energy-density profile and combined with tensor-stability requirements to determine the physically viable parameter region. We further investigate the tensor sector, demonstrating the absence of tachyonic modes through the factorization of the associated Schrodinger-like equation and establishing the localization of the graviton zero mode. Finally, the massive spectrum is analyzed via the relative-probability method, revealing the emergence of odd-parity gravitational resonances associated with the split-brane structure, whose quasi-localization is significantly enhanced near the lower stability boundary ($q\to -1$).

\section{Analytical construction of thick branes in $f(T,T_G)$ gravity}

We consider a five-dimensional modified teleparallel gravity theory, in which the spacetime metric is related to the fünfbein field $e^{a}{}_{M}$ through
\begin{equation}
g_{MN}=\eta_{ab}\,e^{a}{}_{M}e^{b}{}_{N}.
\end{equation}
Adopting the Weitzenböck gauge, the gravitational dynamics are entirely encoded in the torsion scalar $T$ and in the teleparallel Gauss--Bonnet invariant $T_G$. The corresponding action is given by 
\begin{equation}
S=\int d^{5}x\,e\left[\frac{1}{4}f(T,T_G)+\mathcal{L}_m\right],
\end{equation}
where $e=\sqrt{-g}$ and $\mathcal{L}_m$ denotes the matter Lagrangian. The field equations follow from the variation of the action with respect to the fünfbein. Since the explicit forms of $T$, $T_G$, and the associated equations of motion are well established in the literature, we refer the reader to Ref.~\cite{Kofinas2014} for their complete derivation.

We consider a five-dimensional warped spacetime described by  \cite{Moreira:2021uod}
\begin{equation}
    ds^2=e^{2A(y)}\eta_{\mu\nu}dx^\mu dx^\nu+dy^2,
    \label{eq:metric}
\end{equation}
where $A(y)$ is the warp function and $y$ denotes the extra dimension. For this geometry, the torsion scalar and the teleparallel Gauss--Bonnet invariant are $ T=-12A'^2$, 
and $ T_G=96A'^2A''+120A'^4$.

We focus on the linear teleparallel Gauss--Bonnet model \cite{Kofinas2014}
\begin{equation}
    f(T,T_G)=-T+\alpha T_G.
    \label{eq:model}
\end{equation}
Since both $f_T$ and $f_{T_G}$ are constants, derivative terms involving these quantities do not appear in the background equations. 

To generate a smooth thick-brane scenario, we consider a real scalar field
$\phi=\phi(y)$ that depends only on the extra-dimensional coordinate. The matter sector is described by the canonical Lagrangian density
\begin{equation}
\mathcal{L}_{m}
=
-\frac{1}{2}\,g^{MN}\nabla_M\phi\,\nabla_N\phi
-V(\phi),
\end{equation}
where $V(\phi)$ is the self-interaction potential responsible for supporting
the brane configuration.

For a scalar-field source, the effective background equations are
\begin{eqnarray}
    \phi'^2
   &=&
    -\frac{3}{2}A''
    \left(1-4\alpha A'^2\right),
    \label{eq:phi_background}\nonumber\\
    V(y)
    &=&
    -3A'^2
    -\frac{3}{4}A''
    +3\alpha A'^2A''
    +6\alpha A'^4.
\end{eqnarray}

We introduce a superpotential $W(\phi)$ through  \cite{Moreira2021fTT,Bazeia:2024gvy}
\begin{equation}
    A'(y)=-\frac{1}{3}W(\phi).
    \label{eq:superpotential_A}
\end{equation}
Substitution into Eq.~\eqref{eq:phi_background} gives
\begin{eqnarray}
    \phi'
    &=&
    \frac{1}{2}W_\phi
    \left(
        1-\frac{4\alpha W^2}{9}
    \right),\nonumber\\
    V(\phi)
    &=&
    -\frac{1}{3}W^2
    +
    \frac{(9-4\alpha W^2)^2}{648}W_\phi^2
    +
    \frac{2\alpha}{27}W^4,
    \label{eq:potential_W}
\end{eqnarray}
where $W_\phi=dW/d\phi$.

For simplicity, we choose the sine-Gordon  superpotential \cite{Moreira2021fTT}
\begin{eqnarray}
    W(\phi)=3k\sin(\beta\phi),
    \label{eq:sin_superpotential}
\end{eqnarray}
where $k>0$ sets the asymptotic curvature scale and $\beta$ controls the deformation of the scalar sector. From Eq.~\eqref{eq:potential_W}, one obtains
\begin{eqnarray}\nonumber
 \phi(y)&=&\frac{1}{\beta}\arcsin u(y),\\ 
    V(\phi)&=&-3k^2\sin^2(\beta\phi)+6\alpha k^4\sin^4(\beta\phi)\\ \nonumber
    &+&\frac{9}{8}k^2\beta^2\cos^2(\beta\phi)\left[1-4\alpha k^2\sin^2(\beta\phi)\right]^2,
    \label{eq:potential_sin}
\end{eqnarray}
where $u(y)=\sin(\beta\phi)$ satisfies the first-order equation
\begin{equation}
    u'=\frac{3k\beta^2}{2} (1-u^2)(1-q u^2) \qquad \mathrm{with} \qquad q = 4\alpha k^2.
    \label{eq:uprime}
\end{equation}

For $q<0$, writing $q=-s^2$, the implicit solution is
\begin{equation}
    \frac{\operatorname{arctanh}(u) + s \arctan(s u)}{1+s^2}
    = \frac{3k\beta^2}{2}(y-y_0).
    \label{eq:u_implicit}
\end{equation}

Meanwhile, for $0\leq q<1$, one obtains
\begin{equation}
    \frac{\mathrm{arctanh}(u)-\sqrt{q}\,\mathrm{arctanh}(\sqrt{q}u)}{1-q}=\frac{3k\beta^2}{2}(y-y_0).
    \label{eq:u_solution_positive_q}
\end{equation}

Throughout this work, $u(y)$ is obtained by solving Eq.~\eqref{eq:uprime} with the boundary condition $u(0)=0$. Furthermore, we restrict the subsequent analysis to the negative-$q$ branch because it is the physically relevant sector for the emergence of an internal structure in the braneworld. Since $1+|q|u^{2}>0$ throughout the interval $-1<u<1$, the scalar flow remains regular and monotonic, and the field smoothly interpolates between the asymptotic vacua $\phi(\pm\infty)=\pm\frac{\pi}{2\beta}$. Therefore, the matter sector will describe a kink-like configurations centered at $y=0$.

The warp factor is
\begin{equation}
    A(y)
    =
    A_0+
    \frac{1}{3\beta^2(1-q)}
    \ln\left(
        \frac{1-u(y)^2}{1-q\, u(y)^2}
    \right).
    \label{eq:A_u}
\end{equation}
As $u\to\pm1$, one has $A'\to\mp k$, so that
\begin{equation}
    A(y)\sim -k|y|.
\end{equation}
The geometry is therefore asymptotically AdS$_5$ and localized around the brane core.

The energy density is
\begin{eqnarray}
    \rho(y)
    &=&
    k^2
    \Big[
        -3u(y)^2
        +
        \frac{9}{4}\beta^2(1-u(y)^2)(1-q\, u(y)^2)^2\nonumber\\
        &+&
        \frac{3}{2}q\, u(y)^4
    \Big].
    \label{eq:rho_bare}
\end{eqnarray}
At infinity,
\begin{equation}
    \rho_\infty
    =
    k^2\left(-3+\frac{3}{2}q\right).
\end{equation}
We define the shifted density
\begin{equation}
    \Delta\rho(u)=\rho_{\rm bare}(u)-\rho_\infty,
\end{equation}
which isolates the localized brane contribution:
\begin{eqnarray}
    \Delta\rho(u)
    &=&
    \frac{3}{4}k^2(1-u^2)
    \Big[
        3\beta^2q^2u^4
        -6\beta^2q u^2
        +3\beta^2\nonumber\\
        &-&2q u^2
        -2q
        +4
    \Big].
    \label{eq:delta_rho}
\end{eqnarray}
This shifted density satisfies $\Delta\rho(\pm\infty)=0$.

\begin{figure}[htbp]
    \centering
    \includegraphics[width=0.49\textwidth]{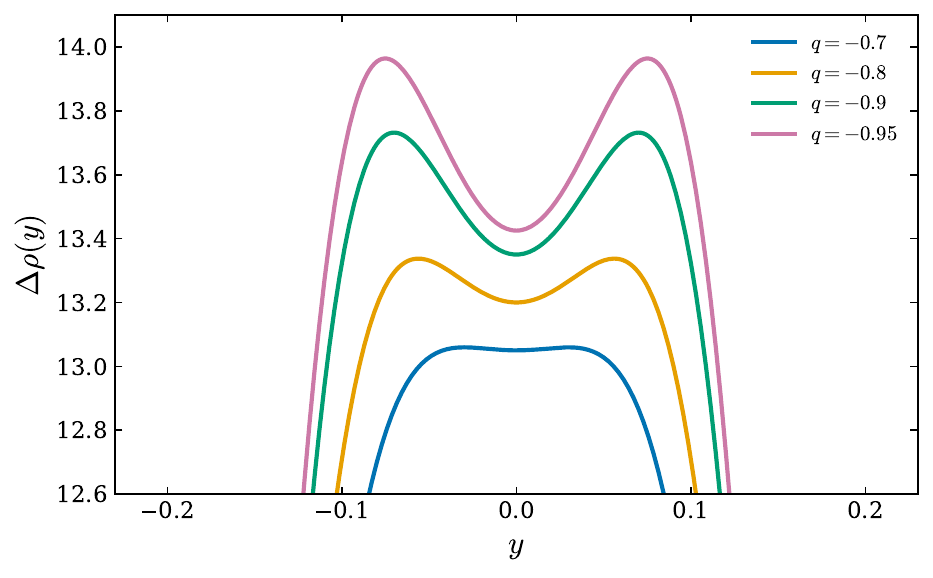}
    \vspace{-1cm}
    \caption{Shifted energy density $\Delta\rho(y)$ for different values of $q$, with $\beta=2$ and $k=1$.}
    \label{fig:delta_rho}
\end{figure}

Expanding around the brane center,
\begin{equation}
    \Delta\rho(u)=\Delta\rho(0)+c_2u^2+\mathcal{O}(u^4),
\end{equation}
one finds that the center ceases to be a local maximum when
\begin{equation}
    q<q_c
    =
    -\frac{1}{2}
    -
    \frac{2}{3\beta^2}.
    \label{eq:qcrit}
\end{equation}
Thus, sufficiently negative values of $q$ induce an internal brane structure leading to brane splitting. This behavior is evidenced in Fig.\ref{fig:delta_rho}.

\section{Tensor perturbations and stable splitting window}

Since the model is linear in both $T$ and $T_G$, it can be related to the curvature formulation through
\begin{equation}
    \bar R=-T+\nabla_M B^M,
    \qquad
    \bar G=-T_G+\nabla_M B_G^M,
\end{equation}
where $\bar R$ and $\bar G$ are the Levi--Civita Ricci scalar and Gauss--Bonnet invariant. Hence,
\begin{equation}
    -T+\alpha T_G
    =
    \bar R-\alpha\bar G+\nabla_M(\cdots).
\end{equation}
Under the usual boundary conditions, the tensor sector of the linear teleparallel model is equivalent to the tensor sector of Einstein--Gauss--Bonnet gravity with
\begin{equation}
    \lambda_{\rm GB}=-\alpha.
    \label{eq:lambda_GB}
\end{equation}
Since $q=4\alpha k^2$, one has
\begin{equation}
    4\lambda_{\rm GB}k^2=-q.
\end{equation}

We consider transverse-traceless tensor perturbations,
\begin{equation}
    ds^2
    =
    e^{2A(y)}
    \left[
        \eta_{\mu\nu}+h_{\mu\nu}(x,y)
    \right]dx^\mu dx^\nu
    +dy^2,
\end{equation}
with
    $\partial^\mu h_{\mu\nu}=0$,
    $h^\mu{}_\mu=0$.
The scalar perturbation is absent in the tensor sector, $\delta\phi=0$.

Writing the tensor fluctuation as a four-dimensional massive mode with extra-dimensional profile $\psi(y)$, the tensor equation takes the Sturm--Liouville form
\begin{equation}
    a(y)\psi''+b(y)\psi'+m^2c(y)\psi=0,
    \label{eq:tensor_abc}
\end{equation}
where
\begin{equation}
    a(y)=K_2(y),
    \qquad
    c(y)=e^{-2A(y)}K_1(y),
\end{equation}
and
\begin{equation}
    b(y)=4A'K_2-8\lambda_{\rm GB}A'A''.
\end{equation}
The kinetic factors are
\begin{equation}
    K_2(y)=1-4\lambda_{\rm GB}A'^2,
    \label{eq:K2_general}
\end{equation}
and
\begin{equation}
    K_1(y)=1-4\lambda_{\rm GB}A'^2-4\lambda_{\rm GB}A''.
    \label{eq:K1_general}
\end{equation}

For $q<0$, the condition $K_2>0$ gives $q>-1$. In the same interval, the last term in Eq.~\eqref{eq:K1_general} is positive, so $K_1>0$. Therefore, the regular tensorial window is
\begin{equation}
    -1<q<0.
\end{equation}
Combining this with the splitting condition \eqref{eq:qcrit}, we find the stable brane-splitting region
\begin{equation}
    \beta>\frac{2}{\sqrt{3}},
    \qquad
    -1<q<-\frac{1}{2}-\frac{2}{3\beta^2}.
    \label{eq:stable_window}
\end{equation}
For $\beta=2$, this gives
\begin{equation}
    -1<q<-\frac{2}{3}.
\end{equation}

\begin{figure*}[ht!]
    \centering
    \includegraphics[width=0.75\textwidth]{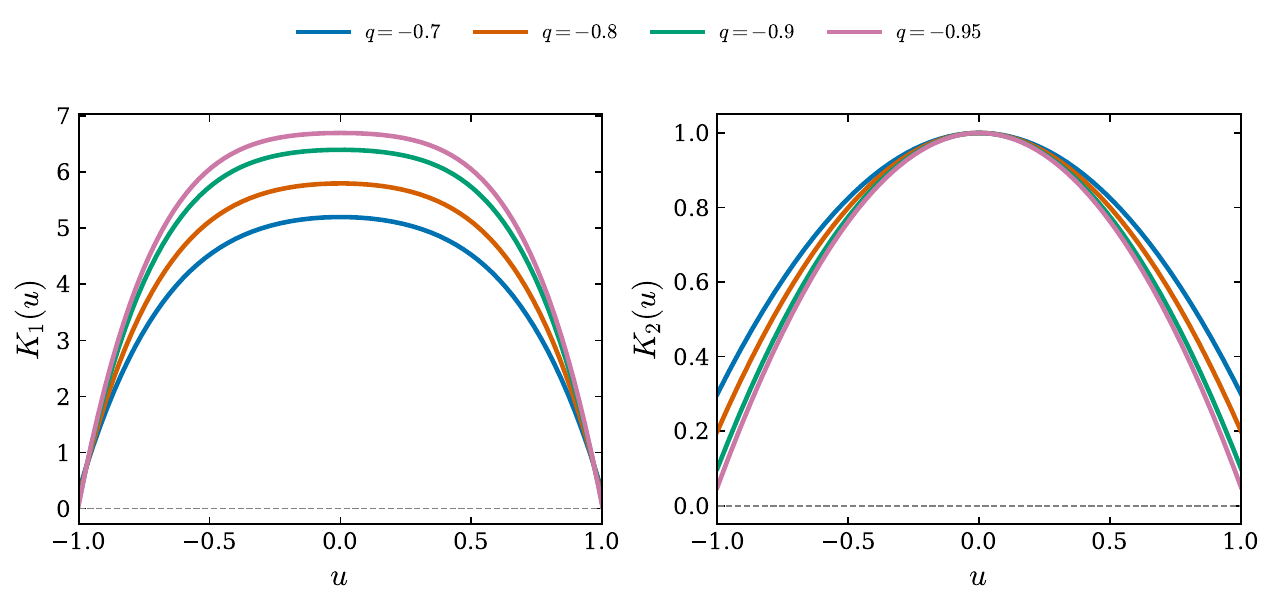}
    \vspace{-0.6cm}
    \caption{Tensorial kinetic factors $K_1(u)$ and $K_2(u)$ for representative values of $q$ inside the stable region.}
    \label{fig:K_factors}
\end{figure*}

Figure~\ref{fig:K_factors} shows that the tensorial kinetic factors \(K_1(u)\) and \(K_2(u)\) remain positive for representative values of \(q\) inside the stable splitting region. This confirms the regularity of the tensor sector and the coexistence of brane splitting with tensor stability.

The tensor equation \eqref{eq:tensor_abc} is transformed into Schr\"odinger form by introducing the Liouville coordinate $w$,
\begin{equation}
    \frac{dw}{dy}
    =
    r(y)
    =
    \sqrt{\frac{c(y)}{a(y)}}
    =
    e^{-A(y)}
    \sqrt{\frac{K_1(y)}{K_2(y)}}.
    \label{eq:liouville_coordinate}
\end{equation}
Using $d/dy=r\,d/dw$, Eq.~\eqref{eq:tensor_abc} becomes
\begin{equation}
    \frac{d^2\psi}{dw^2}
    +
    \mathcal{K}(w)\frac{d\psi}{dw}
    +
    m^2\psi=0,
\end{equation}
where
\begin{equation}
    \mathcal{K}(w)
    =
    \frac{a r'+b r}{c}.
\end{equation}
Removing the first-derivative term through
\begin{equation}
    \psi(w)
    =
    \exp\left[
        -\frac{1}{2}
        \int^w \mathcal{K}(\bar w)d\bar w
    \right]\chi(w),
\end{equation}
one obtains
\begin{equation}
    -\frac{d^2\chi}{dw^2}
    +
    U_{\rm grav}(w)\chi
    =
    m^2\chi.
    \label{eq:schrodinger_tensor}
\end{equation}
The effective gravitational potential is
\begin{equation}
    U_{\rm grav}
    =
    \frac{1}{2}\frac{d\mathcal{K}}{dw}
    +
    \frac{1}{4}\mathcal{K}^2.
    \label{eq:Ugrav_def}
\end{equation}
Defining $\mathcal{J}=\mathcal{K}/2$, we can write
\begin{equation}
    U_{\rm grav}
    =
    \frac{d\mathcal{J}}{dw}
    +
    \mathcal{J}^2.
\end{equation}
Thus,
\begin{equation}
    \left(\frac{d}{dw}+\mathcal{J}
    \right)\left(-\frac{d}{dw}+\mathcal{J}
    \right)\chi=m^2\chi.
    \label{eq:factorized_hamiltonian}
\end{equation}
The Hamiltonian is non-negative whenever $K_1>0$ and $K_2>0$, and therefore tachyonic tensor modes are excluded in the stable region \eqref{eq:stable_window}.

\subsection{Graviton zero mode}

\begin{figure}[ht!]
    \centering
    \includegraphics[width=0.49\textwidth]{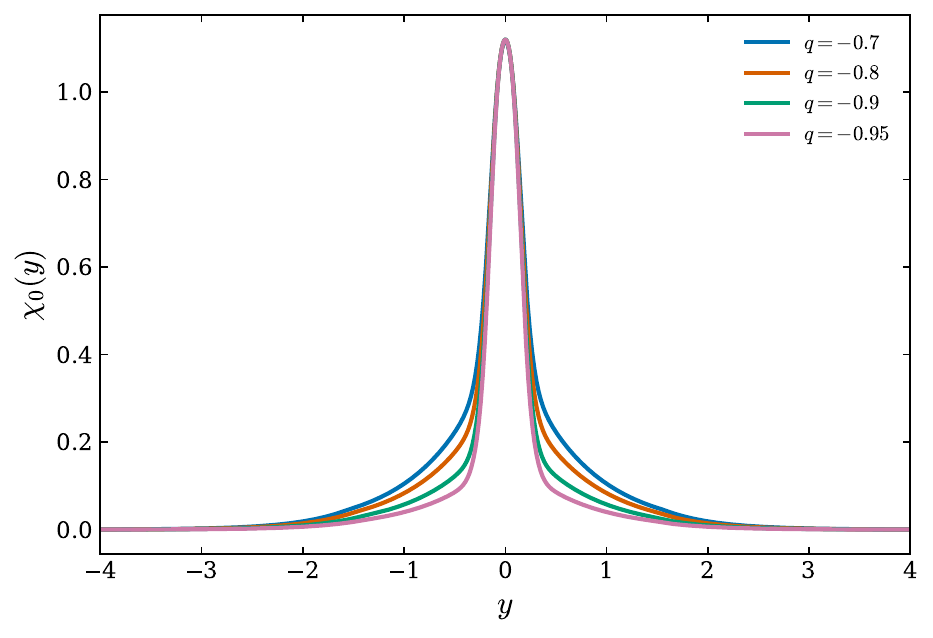}
    \vspace{-1cm}
    \caption{Graviton zero mode $\chi_0(y)$ for representative values of $q$ inside the stable region.}
    \label{fig:chi0}
\end{figure}

The massless graviton mode corresponds to $m=0$. In the original Sturm--Liouville variable, the zero mode is constant, $\psi_0=\text{const.}$ In the Schr\"odinger variable,
\begin{equation}
    \chi_0(w)
    \propto
    \exp\left[
        \frac{1}{2}
        \int^w\mathcal{K}(\bar w)d\bar w
    \right].
\end{equation}
At infinity, $u\to\pm1$, and
\begin{equation}
    K_1\to1+q,
    \qquad
    K_2\to1+q.
\end{equation}
Hence
\begin{equation}
    r(y)
    =
    e^{-A}\sqrt{\frac{K_1}{K_2}}
    \sim e^{-A}.
\end{equation}
Since $A(y)\sim -k|y|$, the normalization integral behaves as
\begin{equation}
    \int |\chi_0(w)|^2dw
    \sim
    \int e^{2A(y)}dy
    \sim
    \int e^{-2k|y|}dy,
\end{equation}
which is finite. Therefore, the zero mode of the graviton is located on the brane as can be seen in Fig.~\ref{fig:chi0}. Furthermore, as \(q\) approaches the lower stability limit \(q\to -1\), the zero mode becomes increasingly concentrated near the center, reflecting the stronger localization induced by the Gauss-Bonnet teleparallel contribution.

\subsection{Massive resonances}

\begin{figure}[ht!]
    \centering
    \includegraphics[width=0.49\textwidth]{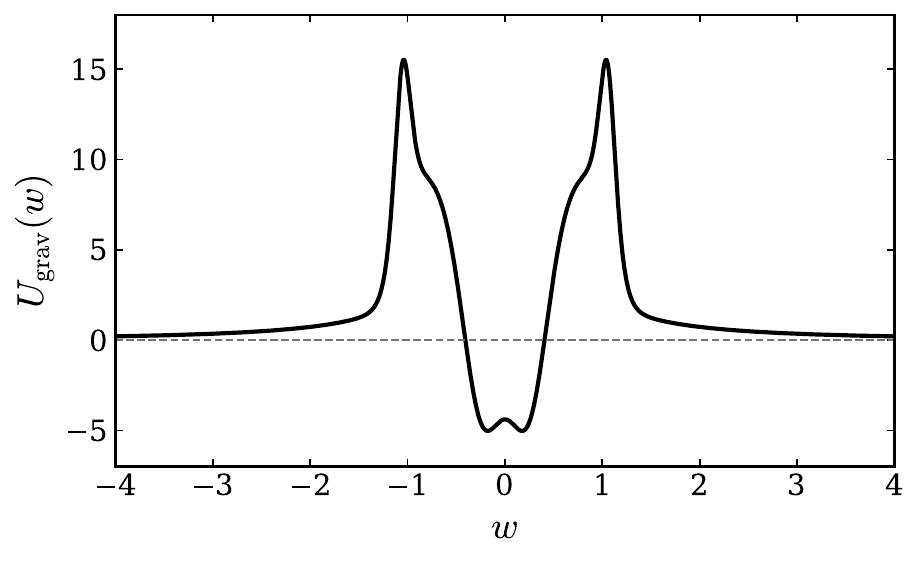}
    \vspace{-1cm}
    \caption{Effective tensor potential $U_{\rm grav}(w)$ for $\beta=2$ and $q=-0.95$.}
    \label{fig:Ugrav}
\end{figure}

To investigate the massive spectrum, we solve
\begin{equation}
    -\frac{d^2\chi_m}{dw^2}
    +
    U_{\rm grav}(w)\chi_m
    =
    m^2\chi_m.
    \label{eq:massive_schrodinger}
\end{equation}
Even and odd parity modes are obtained from
\begin{eqnarray}
    \chi_{\rm even}(0)&=&1,
    \qquad
    \chi'_{\rm even}(0)=0,\nonumber\\
    \chi_{\rm odd}(0)&=&0,
    \qquad
    \chi'_{\rm odd}(0)=1.
\end{eqnarray}

Following the relative-probability method used to identify gravitational resonances in thick-brane models~\cite{Tan2021fTResonances}, we define
\begin{equation}
P(m)
=
\frac{
\int_{-w_b}^{w_b}|\chi_m(w)|^2dw
}
{
\int_{-w_{\max}}^{w_{\max}}|\chi_m(w)|^2dw
}.
\label{eq:relative_probability}
\end{equation}
A peak in $P(m)$ indicates a massive mode with enhanced probability density near the brane. For each resonant peak, $m^2_{\rm res}$ denotes the peak position and $P_{\max}$ its height. The resonance width $\Gamma$ is estimated from the full width at half maximum (FWHM), measured with respect to the local background, and the characteristic lifetime is taken as $\tau\simeq\Gamma^{-1}$ in natural units. Thus, smaller $\Gamma$ corresponds to a longer-lived quasi-localized tensor mode. The prominence $\varpi$ measures the peak height relative to its local background.

For $\beta=2$ and $q=-0.95$ (the effective tensor potential $U_{\rm grav}(w)$ is expressed in Fig.\ref{fig:Ugrav}), the odd sector exhibits a broad peak around
\begin{equation}
    m^2_{\rm res}\simeq 9.9.
\end{equation}
We test its robustness by varying the internal region as
\begin{equation}
    w_b=\frac{w_{\max}}{6},
    \qquad
    w_b=\frac{w_{\max}}{8},
    \qquad
    w_b=\frac{w_{\max}}{10}.
\end{equation}

The peak position remains approximately stable under this variation, whereas its height changes as expected from the definition of (P(m)). No comparable even-parity peak is found in the same mass interval. The extracted parameters are shown in Tab.~\ref{tab:resonance_parameters}; the relatively large value of $\Gamma$ indicates a broad resonance with moderate lifetime.

\vspace{-0.2cm}
\begin{table}[htbp]
    \centering
     \caption{Parameters of the odd tensor resonance for $\beta=2$ and $q=-0.95$.}
    \begin{tabular}{ccccc}
        \hline
        $w_b$ & $m^2_{\rm res}$ & $P_{\max}$ & $\Gamma$ & $\tau\sim 1/\Gamma$ \\
        \hline
        $w_{\max}/6$  & $9.90$ & $0.343770$ & $1.85$ & $0.540541$ \\
        $w_{\max}/8$  & $9.95$ & $0.289403$ & $1.85$ & $0.540541$ \\
        $w_{\max}/10$ & $9.85$ & $0.283480$ & $1.85$ & $0.540541$ \\
        \hline
    \end{tabular}
    \label{tab:resonance_parameters}
\end{table}
\vspace{-0.2cm}

\subsection{Resonance band at $q=-0.999$}

After the global scan, we refined the dominant peaks near
    $q=-0.999$.
The refined data reveal a continuous odd-parity resonance band in the interval
   $ 1.5\leq \beta \leq 3.5$.
Along this band, the resonant mass increases monotonically with $\beta$, while the maximum relative probability grows and saturates close to
   $ P_{\max}\simeq0.88$.
Therefore, the resonant behavior is not associated with a single fine-tuned parameter choice, but with a continuous region close to the lower boundary of the stable domain.

The representative values for $w_b=w_{\max}/8$ are shown in Tab.~\ref{tab:resonant_band_qm0999}. The value of $w_b$ was chosen as an intermediate reference between the narrower and wider integration windows. The same resonance band remains visible for $w_b=w_{\max}/6$ and $w_b=w_{\max}/10$, confirming that the peaks are not artifacts of a particular choice of internal region.

\vspace{-0.2cm}
\begin{table}[htbp]
\centering
\caption{Representative odd tensor resonances for $q=-0.999$ and $w_b=w_{\max}/8$. }
\label{tab:resonant_band_qm0999}
\begin{tabular}{cccccc}
\hline
$\beta$ & $m^2_{\rm res}$ & $P_{\max}$ & $\Gamma$ & $\tau\simeq1/\Gamma$ & $\varpi$ \\
\hline
1.500 & 4.630 & 0.421 & 0.659 & 1.518 & 0.221 \\
1.600 & 5.350 & 0.497 & 0.683 & 1.464 & 0.322 \\
1.700 & 6.185 & 0.593 & 0.610 & 1.639 & 0.453 \\
1.800 & 7.075 & 0.644 & 0.621 & 1.610 & 0.524 \\
1.900 & 7.970 & 0.674 & 0.657 & 1.522 & 0.554 \\
2.000 & 8.915 & 0.713 & 0.654 & 1.529 & 0.582 \\
2.100 & 9.915 & 0.745 & 0.668 & 1.497 & 0.617 \\
2.200 & 10.960 & 0.774 & 0.673 & 1.486 & 0.655 \\
2.300 & 12.055 & 0.799 & 0.677 & 1.476 & 0.677 \\
2.400 & 13.205 & 0.817 & 0.689 & 1.452 & 0.682 \\
2.500 & 14.400 & 0.831 & 0.707 & 1.415 & 0.689 \\
2.600 & 15.640 & 0.841 & 0.729 & 1.371 & 0.703 \\
2.700 & 16.935 & 0.849 & 0.755 & 1.323 & 0.680 \\
2.800 & 18.270 & 0.855 & 0.783 & 1.276 & 0.691 \\
2.900 & 19.660 & 0.860 & 0.810 & 1.235 & 0.686 \\
3.000 & 21.095 & 0.864 & 0.834 & 1.199 & 0.674 \\
3.100 & 22.575 & 0.869 & 0.853 & 1.172 & 0.667 \\
3.200 & 24.105 & 0.873 & 0.869 & 1.150 & 0.657 \\
3.300 & 25.680 & 0.877 & 0.886 & 1.128 & 0.651 \\
3.400 & 27.300 & 0.879 & 0.908 & 1.102 & 0.634 \\
3.500 & 28.975 & 0.881 & 0.933 & 1.072 & 0.626 \\
\hline
\end{tabular}
\end{table}
\vspace{-0.2cm}

The refined resonance band exhibits three main features. First, $m^2_{\rm res}$ grows almost monotonically with $\beta$, showing that the internal-structure parameter controls the position of the massive quasi-localized tensor mode. Second, $P_{\max}$ increases with $\beta$ and saturates near $P_{\max}\simeq0.88$, indicating stronger localization around the brane before reaching a plateau. Third, the width remains moderate,
\begin{equation}
    0.61\lesssim \Gamma \lesssim 0.93,
\end{equation}
with
\begin{equation}
    1.07\lesssim \tau \lesssim 1.64.
\end{equation}
Thus, the modes are not sharply bound states, but broad-to-moderate tensor resonances with enhanced probability density near the brane.

\subsection{Near-boundary refinement at $q=-0.99999$}

The resonance band found at $q=-0.999$ suggests that the quasi-localization of massive tensor modes is enhanced as the model approaches the lower stability boundary $q\to -1$. To test this behavior, we performed an additional near-boundary scan for
    $q=-0.9995$, $q=-0.9999$, $q=-0.99999$,
extending the analysis up to larger values of $\beta$. Since the dominant structure appears in the odd-parity sector, this refinement was restricted to odd modes.

The strongest enhancement occurs for
   $q=-0.99999$.
In this regime, the resonant peaks become substantially higher and narrower than those obtained at $q=-0.999$. The maximum relative probability reaches values above $P_{\max}\simeq0.94$, while the width decreases to approximately $\Gamma\simeq0.50$ in the optimal region.

A fine scan in the interval
    $2.8\leq\beta\leq3.4$,
    $q=-0.99999$,
shows that the optimal region lies around
    $3.0\lesssim\beta\lesssim3.1$.
For the reference internal region $w_b=w_{\max}/6$, the best fine-scan point is located near
    $\beta\simeq3.03$,
with
   $ m^2_{\rm res}\simeq21.02$,
    $P_{\max}\simeq0.946$,
    $\Gamma\simeq0.505$,
    $\tau\simeq1.98$.
Neighboring values of $\beta$ give very similar widths and heights, indicating that this is not an isolated numerical spike, but a stable near-boundary resonant region.

The broad scan up to $m^2=150$ confirms the same behavior. Among the tested points, the strongest long-lived resonance appears around
    $\beta=3.10$,
    $q=-0.99999$,
with
    $m^2_{\rm res}\simeq22.05$,
   $ P_{\max}\simeq0.950$,
    $\Gamma\simeq0.474$,
    $\tau\simeq2.11$,
for $w_b=w_{\max}/6$. The values obtained with $w_b=w_{\max}/8$ and $w_b=w_{\max}/10$ remain close to this result, confirming the robustness of the peak. Representative near-boundary resonances obtained from the broad scan are summarized in Tab.~\ref{tab:near_boundary_best_resonances}, which also illustrates the systematic increase of the relative probability as ($\beta$) grows.

\vspace{-0.2cm}
\begin{table}[htbp]
\centering
\caption{Best near-boundary odd tensor resonances for $q=-0.99999$ from the broad scan up to $m^2=150$.}
\label{tab:near_boundary_best_resonances}
\begin{tabular}{cccccc}
\hline
$\beta$ & $w_b$ & $m^2_{\rm res}$ & $P_{\max}$ & $\Gamma$ & $\tau\simeq1/\Gamma$ \\
\hline
3.00 & $w_{\max}/8$  & 20.60 & 0.9169 & 0.4935 & 2.0262 \\
3.03 & $w_{\max}/6$  & 21.00 & 0.9413 & 0.4943 & 2.0231 \\
3.10 & $w_{\max}/6$  & 22.05 & 0.9495 & 0.4739 & 2.1103 \\
3.10 & $w_{\max}/8$  & 22.05 & 0.9281 & 0.4773 & 2.0951 \\
3.50 & $w_{\max}/6$  & 28.40 & 0.9641 & 0.4889 & 2.0452 \\
3.50 & $w_{\max}/8$  & 28.40 & 0.9545 & 0.4873 & 2.0521 \\
4.00 & $w_{\max}/8$  & 37.40 & 0.9657 & 0.5365 & 1.8638 \\
5.00 & $w_{\max}/8$  & 58.90 & 0.9793 & 0.5746 & 1.7405 \\
6.00 & $w_{\max}/10$ & 85.20 & 0.9802 & 0.7306 & 1.3687 \\
\hline
\end{tabular}
\end{table}
\vspace{-0.2cm}

The near-boundary data show that increasing $\beta$ beyond the optimal region further increases the peak height, with $P_{\max}$ approaching $0.98$ for $\beta=5$ and $\beta=6$. However, this increase in height is accompanied by a larger width, so the lifetime does not improve. Therefore, the best compromise between localization and lifetime occurs for intermediate values of $\beta$, approximately in the range
    $3.0\lesssim\beta\lesssim3.5$.

The broad scan also reveals secondary peaks at larger masses. However, these peaks are considerably wider or have smaller relative probability than the main near-boundary band. For example, secondary structures appear around $m^2\sim30$, $m^2\sim45$, $m^2\sim55$, $m^2\sim90$, and $m^2\sim135$, but they do not provide longer-lived or more sharply localized modes than the main resonance band.

In particular, no ultra-narrow resonance satisfying
    $\Gamma<0.3$,
    $P_{\max}>0.6$
was found in the broad scan up to
    $m^2=150$
for the tested near-boundary configurations. Thus, the model supports robust and highly localized odd tensor resonances near the stability boundary, but the modes remain finite-width resonances rather than ultra-long-lived quasi-bound states.

Combining all scans, the spectral behavior can be summarized as follows. The point $(\beta,q)=(2,-0.95)$ exhibits a broad odd resonance. Moving closer to the stability boundary, the resonance becomes stronger and narrower. At $q=-0.999$, a continuous odd-parity resonance band appears. At $q=-0.99999$, this band develops an optimal near-boundary region around $3.0\lesssim\beta\lesssim3.5$, where the relative probability is high, the width is minimized, and the lifetime reaches $\tau\simeq2.1$. The broad scan confirms that this main near-boundary band dominates the spectrum within the analyzed mass range.

This establishes a direct connection between the internal structure of the brane and the tensor spectrum. The same teleparallel Gauss--Bonnet correction that allows stable brane splitting also generates a tunable band of odd-parity massive gravitational resonances near the stability boundary.

\section{Conclusions}

We have constructed an analytical thick-brane solution in linear teleparallel Gauss--Bonnet gravity, $f(T,T_G)=-T+\alpha T_G$, by means of a first-order formalism generated by a sinusoidal superpotential. The resulting configurations are regular, asymptotically AdS$_5$, and exhibit a controllable internal structure governed by the parameter \(q=4\alpha k^2\). An analytical condition for brane splitting was derived from the shifted energy density and combined with the positivity of the tensor kinetic factors to identify the stable splitting region. The tensor perturbation equation was mapped into a Schr\"odinger-like form with a factorized Hamiltonian, ensuring the absence of tachyonic modes, while the graviton zero mode was shown to be normalizable and localized on the brane.

The massive tensor spectrum reveals a direct connection between teleparallel Gauss--Bonnet-induced brane splitting and gravitational resonances. In particular, odd-parity resonant modes emerge within the stable domain and become increasingly localized as the system approaches the lower stability boundary \(q\rightarrow -1\). A continuous resonance band was identified near this boundary, with the strongest quasi-localization occurring for intermediate values of \(\beta\). These findings demonstrate that teleparallel Gauss-Bonnet corrections provide a simple analytical mechanism for generating stable split branes and tunable massive graviton resonances, highlighting the role of higher-order torsional invariants in the phenomenology of extra-dimensional gravity.

\vspace{2cm}

\section*{ACKNOWLEDGMENTS}
F. C. E. Lima would like to express
their sincere gratitude to the Conselho Nacional de
Desenvolvimento Científico e Tecnológico (CNPq) and
Fundação de Amparo à Pesquisa do Estado de São Paulo
(FAPESP) for their valuable support. F. C. E. Lima
is supported, respectively, for grants No. 2025/05176-7
(FAPESP) and 171048/2023-7 (CNPq).

\end{document}